# "Narco" Emotions: Affect and Desensitization in Social Media during the Mexican Drug War

**Munmun De Choudhury**[1]    **Andrés Monroy-Hernández**[1]    **Gloria Mark**[2]

[1] Microsoft Research
One Microsoft Way
Redmond, WA 98052 USA
{munmund, andresmh}@microsoft.com

[2] Department of Informatics
University of California, Irvine
Irvine, CA 92697 USA
gmark@uci.edu

### ABSTRACT
Social media platforms have emerged as prominent information sharing ecosystems in the context of a variety of recent crises, ranging from mass emergencies, to wars and political conflicts. We study affective responses in social media and how they might indicate desensitization to violence experienced in communities embroiled in an armed conflict. Specifically, we examine three established affect measures: negative affect, activation, and dominance as observed on Twitter in relation to a number of statistics on protracted violence in four major cities afflicted by the Mexican Drug War. During a two year period (Aug 2010-Dec 2012), while violence was on the rise in these regions, our findings show a decline in negative emotional expression as well as a rise in emotional arousal and dominance in Twitter posts: aspects known to be psychological markers of desensitization. We discuss the implications of our work for behavioral health, facilitating rehabilitation efforts in communities enmeshed in an acute and persistent urban warfare, and the impact on civic engagement.

### ACM Classification Keywords
H5.3. Group and Organization Interfaces; Asynchronous interaction; Web-based interaction.

### INTRODUCTION
People living in regions facing warfare or political conflict are often exposed to protracted violence that have negative mental and physical side-effects. For instance, research has shown that prolonged exposure to violence, through media reports, direct observation, word-of-mouth or victimization can bear detrimental impacts on people's affective, and behavioral processes [1]. It can also lead to emotional habituation or numbing, known as "desensitization." Affective desensitization has implications for mental health as it can lead to cognitive performance decline, attentional impairment [20], and is a stressor of the onset of PTSD (post-traumatic stress disorder), an anxiety disorder associated with harmful physiological outcomes [30].

The Mexican Drug War is an example of the type of armed conflict that has exposed people to persistent acts of violence. Since the war started in, many Mexican cities have seen a rapid increase of shootings and homicides that, on occasions, affect innocent civilians. Furthermore, the conflict has triggered an increase of criminal activities such as extortions, and kidnappings affecting the general population [24]. This generalized violence in some Mexican cities, coupled with constrained information reporting on news media, have contributed to the emergence of citizen alert networks using platforms like Twitter and Facebook to inform and collectively grieve, critique, and express frustration about the violence in the streets [25].

Previous research in crisis informatics has demonstrated the role of social media as a lens to understand how society copes with crises and how communities leverage these tools for civic engagement and social support [21,39,27,34,35]. In this paper, we use social media to examine the evolution of affective reactions to persistent violence, and to study whether affective desensitization may be manifested on social media, focusing on the Mexican Drug War. As a first step to this investigation, we focus particularly on Twitter, because of its prominent adoption in the Mexican population—35% of Mexicans are online, of which 82% use social media, and 58% of social media users use Twitter [38].

Given the politically complex and highly contested nature of conflict, we first utilize a multi-prong approach to gauge the context of violence in four major cities in Mexico during Aug 2010 and Dec 2012 (Monterrey, Reynosa, Saltillo, Veracruz). The data sources include: government statistics on homicides, web search interest in violence-related phrases, and social media data from Twitter and a highly popular anonymously curated "narco" blog. Thereafter, we define psychological measures to operationalize affective responses and possible desensitization in Twitter postings of individuals impacted by the conflict: *Negative Affect, activation or arousal, and dominance*.

Our findings show that, while violence was consistent, or in some cases on the rise in the cities we study, Twitter posts discussing the cities indicate a numbing of emotional responses and aggressive expression over time—both of which are known attributes of affective desensitization [6]. Specifically, negative affect shows decline over time, while activation and dominance measures show an increase.

As our data is observational, a causal relationship cannot be inferred; however the results *do* suggest a significant link between exposure to violence due to the ongoing urban warfare in Mexico, and anxiety and post-traumatic stress symptomatology gleaned from social media. Our analyses also contribute to the literature on crisis informatics in general and violence in Mexico in particular, and extend beyond looking at determinants of violence and violence intensity, into studying its consequences manifested on social media. We also discuss the possible implications of our research for civic media.

## RELATED WORK

There has been an increased interest in HCI research in understanding how citizens use social technologies to respond to crises: from natural ones: earthquakes and floods, to manmade ones: wars and terrorist attacks [9,26]. As social media becomes a prominent communication channel, the motivation to study the use of these platforms during crises, comes in part, from a desire to understand how society copes with those events. Our work builds on this prior literature in crisis informatics by investigating how social media can be used to gauge the well-being of a society experiencing prolonged violence. There is limited research in this area e.g., [21] that investigate citizens' affective responses to crisis, even specifically to the Mexican Drug War [26], albeit focused on a US border town. However, to our knowledge, the crisis informatics literature, despite studying a wide range of crises (e.g., [9,26]), has not examined emotional expression of citizens experiencing urban warfare. Our study will investigate emotional responses manifested in social media during a crisis and has implications for public health officials, as social media could potentially be used as a barometer of negative psychological impact on citizens.

However, although not related to crisis informatics, recent research on social media use has demonstrated that it can reveal a variety of behavioral and affective trends. Kramer [19] utilized posts made on Facebook to determine a measure of societal happiness; whereas Golder and Macy [14] found that positive and negative affect expressed on Twitter can replicate known diurnal and seasonal behavioral patterns across cultures. Affect and behavior mined from Facebook and Twitter posts has also been known to be reflective of behavioral [13] and public health concerns [28].

Another thread of research relevant to our work comes from psychology, where the impact of trauma and crisis in the society on the affective responses of people has been widely investigated [22,40]. A critical consequence of a crisis experience is affective desensitization, a well-studied topic in behavioral health domains. Researchers have studied how individuals chronically exposed to violence might experience a decrease in affective and even physical reactions [7]. This phenomenon, affective desensitization, is characterized by the numbing of (negative) emotional reactions to events that typically would elicit a strong response, as well as by an increase in aggressive behavior [6,18]. Additionally, research has shown that even brief exposure to violence both in the physical world [11] and through the media (from television, to video games, and the Internet), can result in affective desensitization [6,7].

The rich body of studies on affect due to trauma, and associated desensitization has all been done in laboratory settings, or with surveys, with homogeneous samples. To our knowledge, they have never been investigated with social media. Social media provides an opportunity to study whether desensitization to violence can be detected on a macro scale by studying the affective responses of people as a consequence of trauma. Due to its archival record, reactions to societal violence can be tracked longitudinally.

Weaving together these observations from prior work, along with the known unique context of social media use in Mexico as a mechanism of expression during the ongoing armed conflict [25] provides us an opportunity to study the affective responses of people enmeshed in the chronic violence, and the possible manifestation of desensitization on Twitter.

## BACKGROUND AND RESEARCH QUESTIONS

### Drug War in Mexico: An Overview

The Mexican Drug War is an ongoing armed conflict among rival drug cartels fighting each other for regional control and against the Mexican government forces and civilian vigilante groups. It was reported in 2011 that this Drug War had taken 60,000 lives [4] and has displaced between 230,000 and 1.6 million people. However unconfirmed reports set the homicide statistics over 100,000 victims that include cartel members (~90% [16]), law enforcement personnel, officials, journalists, and innocent civilians. Beyond mere body counts, these numbers over time are taking a hidden psychological toll on citizens as well: forced disappearances of people, for instance, take a huge emotional toll on the families and the community [23]. Furthermore, small business owners are extorted and threatened with death by petty criminals taking advantage of the climate of fear [8].

Given the acuteness of the circumstances, and as a consequence of weakened and censored traditional media along with the failure of the local governments in appropriate public communication [25], blogs and social media accounts [15] have emerged that attempt to report gory details, warnings, and alerts about the ongoing violence in various cities as they unfold—increasing the average civilian's exposure to the crisis.

The hijacking of the control over violence has also led to a continuing deterioration of the Mexican social fabric, and has been known to lead to detrimental impact on behavioral health, apparent in symptoms of post-traumatic stress and anxiety. In fact, as early as 2010, local health officials had reported a significant increase in the number of people seeking mental health help with post-traumatic stress disorder (PTSD) induced by drug-related violence [26]. Similarly, the international news media have reported how Mexicans are "numb to carnage" [2] and even kids are "exposed to such violence that they're desensitized" [17]. These observations motivate our work to study the affective responses of people manifested in social media in the light of the ongoing armed conflict, and also investigate whether affective desensitization may be present.

**Research Questions**

In the context of the ongoing urban warfare in Mexico and the context described above, we aim to investigate the following core question: *can social media signal how people's affective reactions to violence change as they get exposed to persistent violence?* In the following paragraphs we will flesh out our research questions.

First, in order to characterize affective reactions via people's social media postings, we rely on the following measures:

1. *Negative Affect (NA)*. A broad emotional dimension, that indicates the level of displeasure of an emotion; e.g., *sad* has a higher measure of NA than *bored*.
2. *Activation*. An established dimension of affect that measures the intensity of an emotion, or arousal [12]. As an example, while *frustrated* and *infuriated* are both negative emotions, *infuriated* is higher in activation.
3. *Dominance*. Another well-studied dimension of affect, that represents the "controlling power" of an emotion [12]. For instance, while both *fear* and *anger* are negative emotions, *anger* is a dominant emotion, while *fear* is a submissive emotion.

We now discuss the relationship of these affective measures to desensitization. Studies have indicated desensitization to be associated with the attenuation or elimination of cognitive, emotional, and, ultimately, behavioral responses to a stimulus [31]. Hence operationally, desensitization manifests itself with numbing (or lowering) of negative emotions, i.e., failure to respond appropriately to the anticipated negative consequences of a violent situation. This motivates our first RQ:

> **RQ 1:** *How does the expression of NA in social media change over time, subject to prolonged violence? In particular, what kind of changes in NA would characterize affective desensitization on social media?*

Observing violence may additionally create or accentuate feelings associated with aggressive thoughts [31]. People may thus habituate emotionally to the display of aggression, although their general fear of crime and violence may become enhanced. These aggressive thoughts can often manifest in the form of increased emotional arousal, and the expression of more dominating feelings [11]. We address these notions in our second and third RQs:

> **RQ 2:** *How does the expression of activation in social media change over time, subject to prolonged violence exposure? In particular, what kind of changes in activation would characterize affective desensitization on social media?*

> **RQ 3:** *How does the expression of dominance in social media change over time, subject to prolonged violence exposure? In particular, what kind of changes in dominance would characterize affective desensitization on social media?*

Finally, literature in crisis informatics [21,27,34,35], and research on the drug war-related violence in Mexico [25] has demonstrated that given such extraordinary circumstances, people are likely to use social media (e.g., Twitter) to express frustration, grievances, or simply to share instantaneous information to alert fellow citizens about violent clashes, or to seek and provide social support. Such community practices are often driven by social norms (e.g., helping each other, seeking and providing support) or as an objective way of ensuring safety in one's surroundings amid ongoing warfare. Cognizant of these observations, it is likely that the level of people's social media activity during such situations would not change with any desensitization. We present our final RQ:

> **RQ 4:** *How does the level of social media use change over time, subject to prolonged violence exposure? In particular, what kind of changes in social media activity would characterize affective desensitization on the platform?*

**DATA AND METHODS**

Our data collection proceeded in two parts: Twitter data that we used to infer affective responses referring to four Mexican cities, and statistics on ongoing drug war violence.

| Twitter post (Spanish) | Translated in English |
|---|---|
| lamenta edil de san pedro plagio y homicidio de empresario <url> #monterrey | mayor of san pedro regrets kidnapping and murder of businessman <url>#monterrey |
| voy viendo lo de los ahorcados en #saltillo estoy en shock! en que momento?? gente cuidense mucho y tomen precauciones!!! | I just saw that people were hanged in #saltillo I am in shock! when was that? be careful and take precautions! |
| que bueno que hayan detenido a este hombre que destazaba hombre para hacerlos pasar por carne y venderla #reynosa <url> | It is good that they have arrested this man was butchering people to make them pass by meat and sell it #reynosa <url> |

Table 1. Example city-specific postings on Twitter.

**Social Media Data**

We utilized Twitter's Firehose stream, made available to us via a contract with Twitter. We collected all of the Spanish postings with hashtagged mentions (e.g., #cityname) of one of the four Mexican cities of interest: Monterrey, Reynosa, Saltillo, and Veracruz, between Aug 2010 and Dec 2012 *(retweets were disregarded)*. Example posts are given in Table 1. We did experiment with focusing on only geocoded Twitter posts to accurately collect postings from the residents of the four cities. However, the proportion of geocoded posts was extremely low (under 1%), and hence less likely to be representative of the affective state of the social media users in the cities in our investigation. City hashtags would likely capture the general affective state in a city, as they were explicitly marked to refer to the city, and therefore more appropriate to use to characterize affective responses than drug war specific hashtags, like #mtyfollow, used in [25]. Statistics of the posts per city are given in Table 2.

|  | Monterrey | Reynosa | Saltillo | Veracruz |
|---|---|---|---|---|
| tweets | 672,310 | 501,313 | 608,936 | 1,336,478 |
| users | 91,044 | 19,068 | 28,267 | 81,589 |
| tweets/user | 7 | 26 | 22 | 16 |
| tweets/day | 778.14 | 580.22 | 704.79 | 1546.85 |

**Table 2. Statistics of Twitter data on the four cities.**

**Violence Data**

*Data on homicides by the Government of Mexico*
A concrete proxy of violence in a location, although often a conservative estimate, is the official number of homicides that are reported over a period of time at that location. We referred to the data gathered from the National Institute of Statistics and Geography (*www.diegovalle.net/projects.html#url=%23datasets*) to obtain monthly aggregates of the number of homicides in each of the four cities we focus on, affected by the drug wars from Aug 2010 to Dec 2011 (Monterrey, Reynosa, Saltillo, Veracruz). Since this source did not have statistics from 2012, we referred to the website of the "Executive Secretariat of the National Public Security System" (*www.secretariadoejecutivo snsp.gob.mx*) to obtain comparable monthly aggregates of homicides for 2012.

*Google Search Trends on Drug War Violence in Mexico*
We conjecture that an objective measure of the degree of violence in a location could also be the tendency of individuals to look for information and news on violent events at the same location. Hence we referred to Google Trends (*www.google.com/trends*) to obtain "search interest" measures in monthly averages over a variety of violence related phrases that we handcrafted specific to the drug war context in Mexico (we chose "Spanish" as the default language). Specifically, we used an iterative approach by starting with 10 terms from the "death" category of the prominent psycholinguistic lexicon: Spanish Linguistic Inquiry and Word Count or LIWC (*www.liwc.net*), and then expanded this term set using co-searched terms suggested by Google and one of the researcher's familiarity with the drug war context in Mexico. These phrases included (all searches also included the word "mexico" for additional specificity): *asesinato, cadaver, calcinado, falleció, fosa\*, homicidio\*, incinerar\*, luto, matanza, matar, morir, muere*.

*Postings on Blog del Narco (BDN)*
A crucial aspect of the drug war context in Mexico was that much of Mexico's mainstream media, especially television stations and local newspapers, had shied away from covering killings and naming the cartels involved. This is because Mexico has become one of the world's most dangerous countries for journalists [10]. Between 2005 and 2010 ~ 66 reporters were killed, with 12 more disappearing, according to the National Human Rights Commission (NHRC).

Given the constraints of statistics reported in traditional media, we therefore resorted to unofficial (citizen) reporting of drug war violence; of these, the Blog del Narco (BDN) (*www.blogdelnarco.com*) edited by an anonymous curator is the most notable, has gained much popularity over the years, and has also been a subject of controversy. This blog publishes graphic details of spiraling violence in different Mexican cities [3]. As of September 2013, the number of followers of the blog on its Twitter account (@InfoNarco) was ~135K. Also, according to a book by the blogger herself, in 2012, the blog "recorded an average of 25 million monthly hits and was ranked among the 100 most-visited websites in Mexico by Alexa, and in the 4,000 for the world" [3].

Using a web crawler we collected all of the postings on BDN in our period of interest: Aug 2010 to Dec 2012. We collected the text of all of the post titles, the post itself and the timestamp of its posting. We calculated the number of postings in monthly aggregates to obtain a time series of the volume of the blog's postings. The total number of posts was 6,880 with an average of 7.9 postings per day during our period of study. Additionally, we obtained monthly snapshots of #followers of BDN's Twitter account over the same period of analysis. Twitter's Firehose stream was again used for the purpose. We conjecture that the audience size of the blog on social media (as indicated by the number of Twitter followers) gives a notion of its visibility and popularity in the greater violence-affected community.

*Usage of Narco Language*
The persistent drug war related violence in Mexico saw the emergence of a terminology that was attributed specifically to drug cartel informants, and to describing the drug war victims. While it initially emerged in colloquial speech and news media, it slowly and steadily made its way into social media like Twitter as well. For example: "*encobijado*", "*encajuelado*", "*encintado*", are names given to those killed in drug-related violence based on how their body was found. *Encintado* refers to a body found suffocated in packing tape; *encobijado* is wrapped in a blanket, and *encajuelado* is stuffed in a car trunk.

| narcotunel* | levantado* | capo* | ejecuci* |
|---|---|---|---|
| narcotúnel* | narcomanta* | sicari* | ejecuta* |
| encobijad* | narcomensaje* | decapit* | balea* |
| encajuelad* | narcocorrido* | degolla* | plagi* |
| levanton | narcoejecucion* | destaza* | balacera* |
| levantón | narcoejecución* | ahorca* | |

**Table 3. List of narco words.**

Our motivation was to leverage the volume of usage of this colloquial "narco language" on Twitter into gauging, as another complementary signal, the expression and reporting of drug war-related violent events in our four cities of interest and during the time period of this study. We first created an ensemble of "narco words" by combining information reported in local and national press in Mexico, and lists of words published by the Royal Spanish Academy (official arbitrator of the Spanish language) [37]. A list of these words is given in Table 3.

In order to obtain the volume of usage of these words on Twitter, we utilized the ensemble of posts we collected over the four cities, and followed a regular expression matching exercise on the content of the post to determine if it mentioned a particular 'narco' word. Like other data sources, we constructed monthly aggregates thereafter by computing the mean proportion of posts in each month that contained one or more narco words. We obtained approximately 14.5K narco word-containing posts in the case of Monterrey, 17.9K in the case of Reynosa, 16.6K for Saltillo, while 20.8K for Veracruz. Example Twitter posts mentioning popular narco words are given in Table 4.

In the following sections, we utilize these data sources to first establish the persistent nature of violence in our regions of interest, and then examine the four RQs we set up in the early part of the paper.

| Twitter post (Spanish) | Translated in English |
|---|---|
| *jornada violenta ayer en #monterrey deja saldo de 5 ejecutados <url> @milenio* | *violent day yesterday in #monterrey leaves 5 executions <url> @milenio* |
| *@trackcoahuila delincuentes se dan a la fuga tras breve balacera con militares en v. carranza y figueroa en #saltillo // esta confirmado???* | *@trackcoahuila criminals on the run after a brief shootout with soldiers in v. Carranza and Figueroa on # saltillo // is this confirmed???* |
| *ejecutan a mujer, pkm ya con estas van 80 ejecutadas, malditos narcos #verfollow #veracruz #pozarica* | *woman executed, wtf there's been 80 women execute, damn narcos # verfollow #veracruz #pozarica* |

Table 4. Sample posts showing use of narco words.

## PERSISTENCE OF DRUG WAR VIOLENCE IN MEXICO

We first begin by discussing how our sources of data convey the persistence of drug war related violence in Mexico, beyond what is indicated by government reported homicide statistics. The purpose of this analysis is to establish factual evidence that demonstrates the ongoing nature of drug war violence in Mexico in general, and the four cities in particular during our period of analysis.

### Violence-related Search Interest, BDN Statistics

The web search data from Google Trends (Figure 1) show that search interest specific to the violence keywords has a stable trend throughout—notice the linear trend fit. Also, we notice that the number of postings on BDN and its number of Twitter followers (Figure 2) show a slight increase (the positive slope of the linear trend fit). Specifically, for search interest score the mean values are 34.09 in 2010-11 and 35.61 in 2012 (little change; $p=.2$ based on independent sample $t$-test). The same for the number of posts on BDN is an average of 227.88 in 2010-11 and 240.57 in 2012 which indicates a slight increase ($p<.05$ based on independent sample $t$-test). In the case of the Twitter follower count of the blog as well, we observe the mean #followers in 2012 (=109,698) to be greater than that in 2010-11 (=54,948) (significant change $p<10^{-5}$, per independent sample $t$-test). Together, these observations imply the presence of persistent drug war related violence during our period of study.

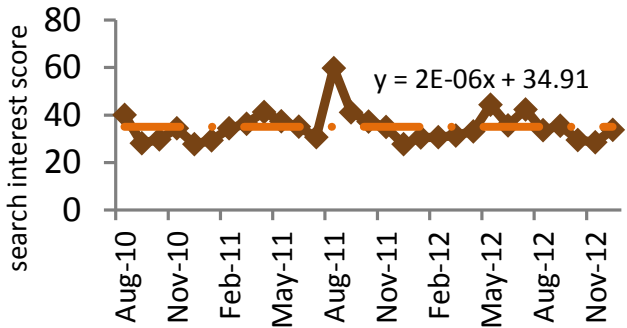

Figure 1. Distributions of search interest scores on violent terms specific to Mexico.

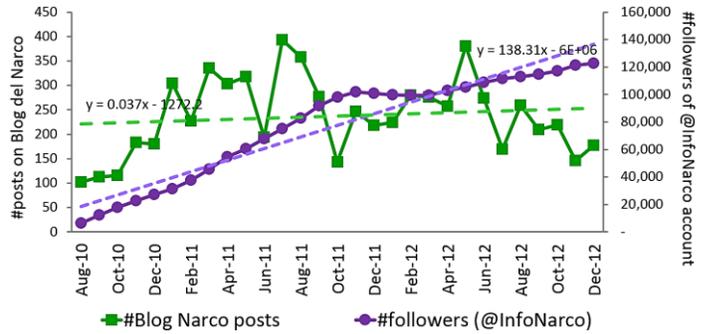

Figure 2. Number of posts on BDN, and number of followers on its Twitter account over time.

How well do these aggregate countrywide statistics match up to the city-specific government reported homicide counts? Compared to the city-specific numbers, we find that the distributions of search interest scores show positive correlations with number of homicides: for Monterrey $r^2=.33$ ($p<10^{-8}$); Reynosa $r^2=.28$ ($p<10^{-15}$); Saltillo $r^2=.31$ ($p<10^{-15}$); and Veracruz $r^2=.42$ ($p<10^{-4}$). In the case of number of postings on BDN, correlations with #homicides are positive as well: for Monterrey $r^2=.57$ ($p<10^{-10}$); for Reynosa $r^2=.34$ ($p<10^{-14}$); for Saltillo $r^2=.41$ ($p<10^{-15}$); and for Veracruz $r^2=.27$ ($p<10^{-14}$). In the same light, positive correlations between the number of homicides in each city and number of BDN's Twitter followers are noted: Monterrey $r^2=.17$ ($p<10^{-11}$); Reynosa $r^2=.13$ ($p<10^{-11}$); Saltillo $r^2=.48$ ($p<10^{-11}$); and Veracruz $r^2=.25$ ($p<10^{-11}$).

In essence, the positive values of the correlation coefficients suggest that search interest scores, number of postings on BDN and its number of Twitter followers additionally support the government data on homicides reflecting the ongoing urban warfare in the cities of interest.

### Narco Language Use

Next, we shift gears to exploring levels of violence in the four cities of interest (Monterrey, Reynosa, Saltillo, and Veracruz). Here we present the volume of Twitter posts corresponding to each city that contain a narco word, between Aug 2010 and Dec 2012 (Figure 3). Each of the four distributions shows an increasing trend, implying that usage of narco-specific language gradually evolved on Twitter in the light of the ongoing urban warfare situation. For instance, the per month mean volumes of narco word-containing posts in 2010-11 are: 464.5 (14%) for Monterrey, 341.4 (22%) for Reynosa, 141.7 (15%) for Saltillo, and 488.9 (28%) for Veracruz. On the other hand, the same numbers in 2012 are higher: 554.2 (18%) for Monterrey, 1005.3 (29%) for Reynosa, 1180.2 (32%) for Saltillo, 1039.5 (30%) for Veracruz. Across the entire period, Pearson correlation coefficient comparing narco word use with number of homicide yields positive values: $r^2=.41$ ($p<10^{-5}$) for Monterrey, $r^2=.29$ ($p<10^{-5}$) for Reynosa, $r^2=.33$ ($p<10^{-4}$) for Saltillo, and $r^2=.46$ ($p<10^{-6}$) for Veracruz. Thus over time, while the number of homicides in the cities were consistent or on the rise, use of narco language increased, providing us additional evidence of ongoing violence.

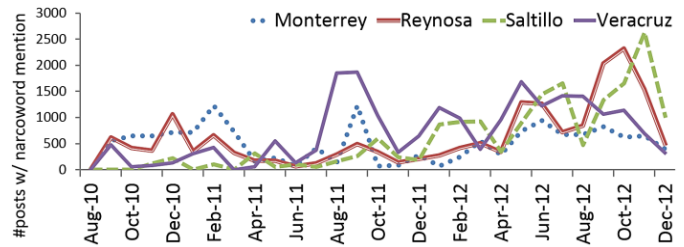

Figure 3. Volumes of Twitter posts containing narco words.

### RESULTS

Having established evidence for persistent violence over the two year period, we now address our research questions.

### RQ 1-3: NA, activation and dominance

**Measuring NA, activation, dominance.** We first discuss our method of computing NA, activation, and dominance values in Twitter posts. First, to compute *NA*, we utilize LIWC (Spanish version), which identifies over 64 behavioral and psychological dimensions (e.g., "insight", "certainty", "perception") in text. LIWC's emotion categories are large in size, broad in semantics, and have been validated for affect computation on Twitter [12,14].

We focus on the words and word stems available in the *negative affect* categories in LIWC: "negative emotion", "sadness", "anger", "anxiety" and "inhibition". We compiled a *negative affect* lexicon for Spanish text. Based on regular expression matching of the words in the lexicon with the content of a Twitter post, we determined a measure of *NA* as the ratio of the number of negative words in the post to the number of words in the post.

|  | decrease in NA in 2012 over 2010-11 | *t*-stat (w.r.t. #homicides) | *p*-value |
|---|---|---|---|
| *Monterrey* | 41.11% | 2.2481 | 1.12e-11 |
| *Reynosa* | 20.76% | 1.1925 | 7.96e-11 |
| *Saltillo* | 12.52% | 1.9845 | 1.66e-11 |
| *Veracruz* | 35.87% | 1.6393 | 0.000197 |

**Table 5. Change in NA in 2012 compared to 2010-11. We also show the results of statistical significance comparing *NA* over time and number of homicides for the corresponding cities over time, using a paired *t*-test (df=29).**

For the other two measures, activation and dominance, we utilized the ANEW lexicon [5] that provides a set of normative emotional ratings for a large number of words in the English language (~2000). For our purpose, we constructed a Spanish resource for the same with the help of translations from two Spanish proficient individuals. To compute activation and dominance, we followed a regular expression match exercise, as with NA measurement. Thereafter, using the corresponding activation (and dominance) values of each such word from ANEW, we determined a mean measure of activation (and dominance) per post, and thereafter computed monthly averages.

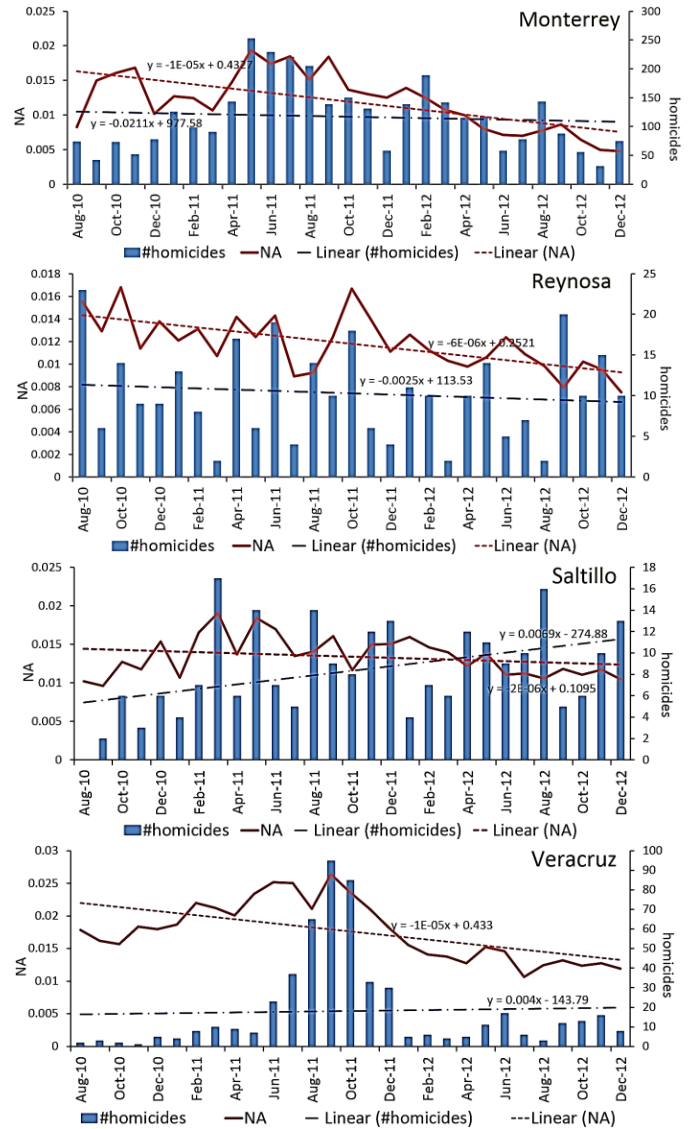

**Figure 4. NA and government reported homicide statistics over time for the four cities.**

**Findings.** Figure 4 shows the trends of *NA* expressed on Twitter over time for each of the four cities: Monterrey, Reynosa, Saltillo, and Veracruz. It also shows the corresponding government reported statistics of homicides, and the respective linear trends for both. From the slope of the trendlines it is clear that, like before, homicides over time show either an increasing trend or a consistently stable trend, and NA for each of the cities shows decreasing trends. Results of a paired *t*-test comparing the trends of the number of homicides (monthly units) with that of *NA* yields statistical significance (see Table 5).

Delving a little deeper, notice that actually towards the beginning of the time period of analysis, i.e., early on in 2010 or early 2011, the peaks in number of homicides are actually correlated with those in NA. However over time, especially in 2012, that ceases to be the case. In Table 5, we observe that in 2012, the mean NA per month was between 12-41% *less* than the mean *NA* per month (i.e., between Aug 2010 and Dec 2011) for all of the four cities. In fact, the Pearson correlation coefficient between the NA distribution and the homicide distribution shows that for each city, in 2010-11 they were highly correlated (Monterrey $r^2$=.81, Reynosa $r^2$=.61, Saltillo $r^2$=.76, Veracruz $r^2$=.63). Whereas in 2012 they followed either an inverse relationship or are uncorrelated: Monterrey $r^2$=.53, Reynosa $r^2$=-.37, Saltillo $r^2$=-.58, Veracruz $r^2$=.07. (We also compare these correlations per city in 2010-11 and in 2012 for significance using Fisher's *r*-to-*z* test. See Table 6).

|  | Monterrey | Reynosa | Saltillo | Veracruz |
| --- | --- | --- | --- | --- |
| **NA** | | | | |
| 2010-11 | 0.81 | 0.61 | 0.76 | 0.63 |
| 2012 | 0.53 | 0.37 | 0.58 | 0.07 |
| *z*-score | 1.26+ | 0.75 | 0.78 | 1.57* |
| **activation** | | | | |
| 2010-11 | 0.84 | 0.41 | 0.52 | 0.83 |
| 2012 | 0.06 | 0.07 | 0.19 | 0.09 |
| *z*-score | 2.72** | 0.86+ | 0.9+ | 2.57** |
| **dominance** | | | | |
| 2010-11 | 0.82 | 0.41 | 0.49 | 0.84 |
| 2012 | 0.41 | 0.13 | 0.08 | 0.16 |
| *z*-score | 1.69* | 0.71 | 1.07+ | 2.48** |

**Table 6.** Fisher's *r*-to-*z* tests, that compare the correlations of measures (NA, activation, dominance) in 2010-11 with 2012. Here + $p < .2$; * $p < .05$; ** $p < .01$.

From the literature, we know that affective desensitization is not an instantaneous phenomenon: rather it happens as a consequence of the convergence of prolonged exposure to real-world and media violence [18]. Along those lines, it appears that the decreasing trend or lowered correlation with number of homicides may be associated with the Twitter users' exposure to the ongoing drug war situation which results in numbing of their negative reactions predominant in the early phases: this could indicate evidence of affective desensitization. *Hence, in the context of RQ 1, lowered NA over time manifested on Twitter may be indicative of affective desensitization.*

Next, in Figure 5 we show activation manifested in the Twitter posts of the four cities over time, along with number of homicides, and their respective linear trend fits. Over time, from the trends of activation, we observe a general increase (ref. the slopes of the linear trend fits for the cities), implying that Twitter users mentioning the four cities in their postings, were increasingly using higher intensity emotions. To elaborate further, we observe that in 2012, the mean activation per month was between 9.9-71% *greater* than the mean activation per month (i.e., between Aug 2010 and Dec 2011) for all of the four cities (Monterrey= 9.9%; Reynosa= 36.7%; Saltillo= 70.8%; Veracruz= 20.5%).

This is also supported by the fact that the Pearson correlation coefficient comparing the activation distribution and homicide distribution gives considerably *lower* (even uncorrelated) values in 2012 than in 2010-11 (ref. Table 6 for details): (2010-11) Monterrey $r^2$=.84, Reynosa $r^2$=.41,

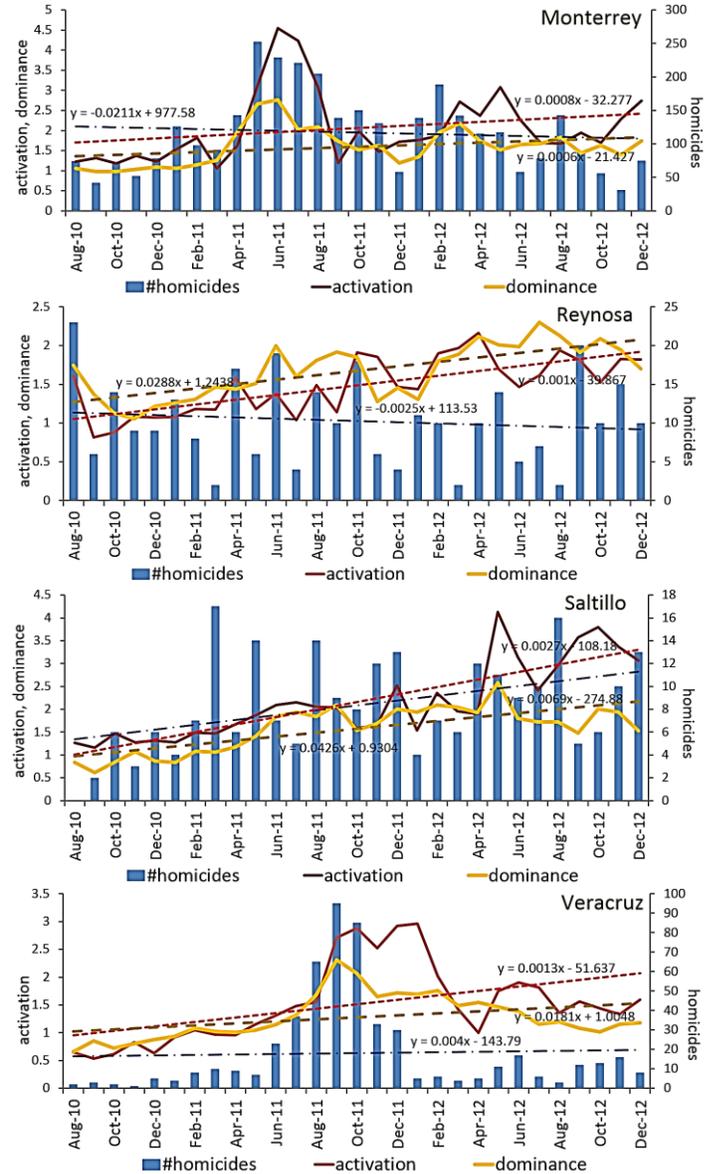

**Figure 5.** Activation and dominance over time for the four cities. Govt. reported homicide statistics are also shown.

Saltillo $r^2$=.52, Veracruz $r^2$=.83; (2012) Monterrey $r^2$=-.06, Reynosa $r^2$=-.07, Saltillo $r^2$=.19, Veracruz $r^2$=-.09. In other words, activation is more reflective of homicide rate early on in our analysis; however it ceases to do so in the later period, indicating possible aggressive tone in Twitter posts [33].

Finally, in Figure 5 we also show the levels of *dominance* in Twitter posts of the four cities over time, along with fit linear trends. Like before, *dominance* shows a rise with persistent violence (ref. the slope of the linear fit), indicating that users are increasingly using dominating and aggressive emotions. We observe that in 2012, the mean *dominance* per month was between 8.6-41.7% *greater* than the mean *dominance* per month (i.e., between Aug 2010 and Dec 2011) for all of the four cities (Monterrey= 8.6%; Reynosa= 29.3%; Saltillo= 41.7%; Veracruz= 9.2%).

Like in the case of the *activation* measure, a Pearson correlation coefficient comparing the *dominance* distribution and homicide distribution gives *lower* (and even uncorrelated) values in 2012 than in 2010-11 (ref. Table 6 for details): (2010-11) Monterrey $r^2$=.82, Reynosa $r^2$=.41, Saltillo $r^2$=.49, Veracruz $r^2$=.84; (2012) Monterrey $r^2$=.41, Reynosa $r^2$=-.13, Saltillo $r^2$=-.08, Veracruz $r^2$=-.16.

To summarize here, rising trends of *activation* and *dominance* over time show signs of aggression in Twitter postings related to the four Mexican cities—a key aspect that characterizes *affective desensitization* [18]. With the persistent number of homicides in the four cities over the time period between 2010 and 2012, it is consistent with the idea that prolonged exposure to violence is associated with desensitization in terms of users' affect expression, with a habituated tendency to display aggression in their day to day postings on social media. *Per RQs 2 and 3 these observations show that affective desensitization on social media may manifest itself via raised levels of activation and the dominance of emotional postings.*

### RQ 4: Levels of social media use

Finally, we present the distributions of the levels of Twitter activity (number of postings) in each of the four cities, for the time period of our analysis. The mean number of posts in monthly aggregates are shown (Figure 6). Like with Figures 1-3, we also show the # homicides per month, and fit linear trend lines for both distributions. The volume of postings for Monterrey and Reynosa show fairly stable trends (ref. slopes of the respective trend fits), whereas Saltillo and Veracruz show an increasing trend. Note that these trends are either considerably highly correlated with the number of homicides—e.g., Monterrey ($r^2$=.46; $p$<10$^{-10}$), Reynosa ($r^2$=.51; $p$<10$^{-20}$), Saltillo ($r^2$=.43; $p$<10$^{-9}$), or show marginal positive correlation—e.g., Veracruz ($r^2$=.002; $p$<10$^{-7}$).

Combining these observations, we conclude that unlike the three affective measures, exposure to violence does not correlate with lowered levels of Twitter activity in the four cities. In fact, as one would expect, given an ongoing urban warfare situation, users resorted to using social media tools over the entire period of our analysis consistently, probably conforming to community practices of providing social support and spreading safety information and alerts to concerned citizens [25]. *Per RQ 4, we therefore observe that social media activity do not show a pattern of a decreasing trend in the context of prolonged exposure to violence.*

### DISCUSSION

Our findings have demonstrated how chronic exposure to violence as a consequence of urban warfare in Mexico is associated with lowered affective responses in Twitter posts of citizens experiencing the violence, leading to possible signs of desensitization in their social media postings. These observations extend the literature on crisis informatics. The Drug War in Mexico is a political conflict that has witnessed weakening of government and traditional media in their ability to keep communities informed of the violence. Twitter has acted as a unique platform allowing affected people to express their emotion, be it their frustration, or grievances and anger about their circumstances anonymously or pseudonymously, amid feelings or perception of threat or terror. This has helped create a counter-narrative to the mainstream media coverage—an aspect that can eventually inform historical accounts about the time period. Our findings can also help researchers build theories about socio-psychological responses to crisis, as well as help compare citizens' thoughts and feelings with their habitual states prior to the upheaval. Additionally, understanding how long certain negative emotional states linger may give an alternate sense of the severity and longevity of a crisis.

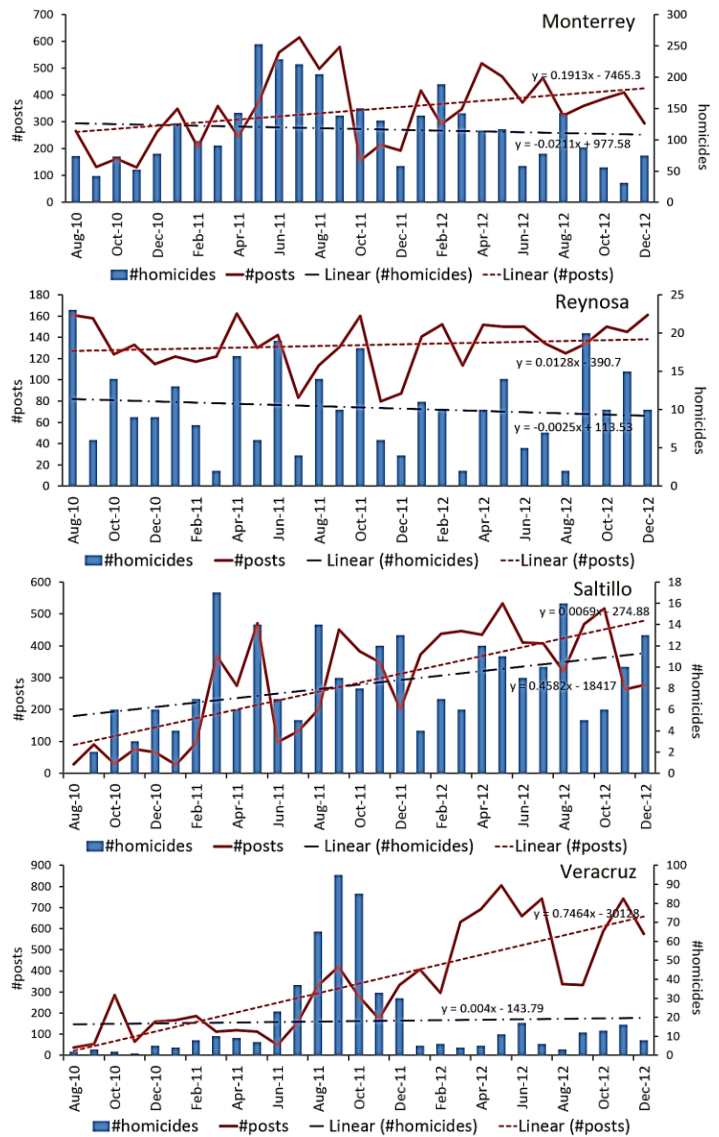

**Figure 6. #posts and homicide stats over time for the four cities.**

### Practical and Design Implications

This research bears implications for public health. Timely attention to communities exposed to violence, experiencing psychological challenges or likely desensitization holds promise in improving rehabilitation efforts in a protracted war context. This is currently a challenge with government or agency administered surveys, because of the large temporal gaps in which they are administered and due to their observations being mostly retrospective. The surveys thus lack the contextualization that is needed to associate the manifested affective concerns to actual circumstances people face. Complementing these surveys with near real-time information about people's affective reactions from social media seems to be an effective way of informing national or international healthcare policies, programs, interventions, and prevention strategies, as well as decisions about the allocation of humanitarian aid.

Our findings also bear implications for the role of civic media during times of prolonged crises. Prior work has indicated that social media has enabled Mexicans to create citizen alert networks, where traditional new media has weakened [25]. While we do not for sure know what people's intrinsic motivations for participation in this case are, we can speculate that strong emotions motivated them to take some risks and use to social media to reach collective action. However, as people experience affective desensitization, it is possible that these collective efforts weaken. More generally, if people rally around common causes when they are passionate about them, then losing that passion could threaten collective action, and perhaps a desensitized society might be less likely to be civically engaged.

As researchers and designers, we therefore need to build information channels that particularly cater to such unique circumstances, and help people gain awareness of the possibility of experiencing *affective desensitization*. Empowering people with tools and platforms that can provide them with psychological and emotional support in the context of widespread affective numbing constitutes a design direction worthy of future exploration.

**Limitations and Future Directions**
Our findings, however need to be interpreted with caution:

**Causality.** First and most importantly, we acknowledge that our study is purely *correlational*. In the absence of ground-truth knowledge about affective desensitization in the society or in social media, we cannot draw causal conclusions regarding whether prolonged violence exposure actually leads to desensitization in people embroiled in the armed conflict. An important future extension would be to collect quantitative evidence of desensitization in the cities affected by the ongoing violence, and examine if our findings, as derived from social media, are predictive of that.

**Alternative explanations.** Certainly, there could be alternative explanations behind our observations. It is possible that with the protracted nature of the warfare ultimately individuals adapt to the environment—a process that may appear as desensitization. Perhaps the society slowly makes its way to normalcy and people try to recover the behaviors they experienced before the violence escalated [21]. However, even if this is the case, we feel that a process of returning to normalcy would involve some degree of desensitization to be able to go about normal day-to-day life. In this light, an analysis of the topics of the social media postings in the cities under consideration, and supplementing with qualitative interviews with affected people may provide insight. It is also possible that people have started to self-censor their Twitter posts out of fear of retaliation—an aspect that may manifest as desensitization.

In fact, desensitization may also be because of exhaustion or long-term stress over time—e.g., simply being wary of the ongoing crisis, instead of being overtly distressed, grieving, or angry. Future research may reveal whether there were corresponding changes in people's daily activities and behavior as well, or whether the affective changes on Twitter were primarily in people's perspectives and thoughts.

**Context.** Finally, Twitter users who share information about a city could be affected by the war directly, through friends and family, indirectly as a bystander, or not affected but simply cognizant about the violent situation. Such distinction might bear a relationship to the degree to which a person shows affective desensitization; however, we do not distinguish between those roles. Investigating these nuances of impact of violence on a person's geosocial context forms interesting directions for future research.

**Social Network, Cultural Effects.** There are also a number of empirically motivated questions around network effects of affective expression and desensitization. Are there network contagion attributes to these phenomena? If so, what kind of attributes of the individual (e.g., personality, gender, age, socio-economic status) that make some more vulnerable to the desensitization process than others? Finally, our work is a case study on one (but notable) ongoing armed conflict in a unique cultural context: in the future it would be interesting to study how our findings generalize to other violence-laden crises in the world, and perhaps other online social platforms.

**CONCLUSION**
Through a large-scale quantitative study around the Mexican Drug War, we have demonstrated the nature of affective changes in social media posts by people exposed to this protracted violence. First, we used official homicide statistics as well as unofficial data (search interest, "narco" language use on Twitter, activity on and visibility of a prominent "narco" blog) to establish factual evidence of ongoing violence in four major Mexican cities in 2010-12. Thereafter, we quantified the affective responses of people in their Twitter posts through the measures *negative affect*, *activation*, *dominance*. While violence was on the rise in our regions of interest, our findings showed a decline in negative affect as well as a rise in emotional arousal and dominance in Twitter posts: aspects that are known to be psychological correlates of desensitization. We discussed how our findings may augment mental health related rehabilitation efforts in crisis afflicted communities, and how civic media might need to adapt to address this psychological challenge.

**ACKNOWLEDGMENT**
This material is partly supported by NSF grant #1218705.